\documentclass[a4paper,10pt,twoside]{cpc-hepnp}

\usepackage{multicol}
\usepackage{graphicx}
\usepackage{booktabs}
\usepackage{amssymb,bm,mathrsfs,bbm,amscd}
\usepackage[tbtags]{amsmath}
\usepackage{lastpage}

\usepackage{graphicx}
\usepackage{xspace}
\usepackage{amsfonts}
\usepackage{srcltx}
\usepackage{hyperref}

\newcommand{\Gee}{\ensuremath{\Gamma_{ee}}\xspace}
\newcommand{\Gll}{\ensuremath{\Gamma_{\ell\ell}}\xspace}
\newcommand{\Bll}{\ensuremath{\mathcal{B}_{\ell\ell}}\xspace}

\newcommand{\GeeBll}{\ensuremath{\Gamma_{ee}\times\Gamma_{\ell\ell}/\,\Gamma}\xspace}
\newcommand{\GBee}{\ensuremath{\Gamma_{ee}\times\Gamma_{ee}/\,\Gamma}\xspace}
\newcommand{\GBmumu}{\ensuremath{\Gamma_{ee}\times\Gamma_{\mu\mu}/\,\Gamma}\xspace}
\renewcommand{\Re}{\ensuremath{\text{Re}}\,}
\renewcommand{\Im}{\ensuremath{\text{Im}}\,}
\renewcommand{\epsilon}{\varepsilon}

\begin{document}

  \fancyhead[co]{\footnotesize 
    A.G. Shamov [KEDR collaboration]: 
    Measurement of $J/\psi$ leptonic width with the KEDR detector}

  \title{Measurement of $J/\psi$ leptonic width with the KEDR detector
    \thanks{Partially supported by the Russian Foundation for Basic
      Research, Grants 08-02-00258, 09-02-08537 and 
       RF Presidential Grant for Sc. Sch. NSh-5655.2008.2
      }
  }

  \author{
      A. G. Shamov$^{1,1)}$\email{A.G.Shamov@inp.nsk.su}
      [KEDR collaboration]
    }
    \maketitle

    \address{
      1~(Budker Institute of Nuclear Physics, 11, 
      Lavrentiev prospect,  Novosibirsk, 630090, Russia)\\
    }

  \begin{abstract}
    We report a new precise determination of the leptonic widths
    of the J/psi meson performed with the KEDR detector at the
    VEPP-4M $e^{+}e^{-}$ collider. The measured values of the
    J/psi parameters are:
    \begin{equation*}
      \begin{split}
        &\GBee\,=0.3323\pm0.0064\,\text{(stat.)}\,\pm0.0048\,\text{(syst.)}\,\,\text{keV},\\ 
        &\GBmumu=0.3318\pm0.0052\,\text{(stat.)}\,\pm0.0063\,\text{(syst.)}\,\,\text{keV}.   
      \end{split}
    \end{equation*}
Assuming $e\mu$ universality and using the table value of the
branching ratios the leptonic
\(\Gamma_{\ell\ell}=5.59\pm0.12\,\text{keV}\) width and the total
\(\Gamma=94.1\pm2.7\,\text{keV}\) widths were obtained.
        We also discuss in detail a method to calculate radiative
        corrections at a narrow resonance.
  \end{abstract}


\begin{multicols}{2}

\section{Introduction}
\label{sec:intro}

The \(J/\psi\) meson is frequently referred to as a hydrogen atom for
QCD.  The electron widths \(\Gamma_{ee}\) of charmonium states are
rather well predicted by potential
models~\cite{Badalian:2008bi,lakhina-2006-74}. The uncertainty in the
QCD lattice calculations of \(\Gamma_{ee}\) gradually approaches the
experimental errors~\cite{dudek-2006-73}.  The full and
dileptonic widths of a hadronic resonance, \(\Gamma\) and
\(\Gamma_{\ell\ell}\), describe fundamental properties of the strong
potential~\cite{brambilla-2005}.

In this report we discuss the results of the \(J/\psi\) meson observation in
leptonic decay channels. Study of the 
$e^{+}e^{-} \to J/\psi \to \ell^{+}\ell^{-}$ cross section
as function of energy allows one to determine
the leptonic width $\Gamma_{{\ell\ell}}$ and its product to the decay ratio
$\Gee\times\Gamma_{\ell\ell}/\Gamma$ thus the total width $\Gamma$ can be
also found. The product $\Gee\times\Gamma_{\ell\ell}/\Gamma$ determines
the peak cross section while the leptonic width $\Gll$ is contained in
the interference wave magnitude. Due to smallness of the interference
effect the experimental accuracy of the $\Gll$ determination is
rather poor. However, the branching ratio \Bll is known with
the accuracy of 0.7\% from the cascade decay 
$\psi(2S)\to J/\psi\,\,\pi^{+}\pi^{-}$ thus we report the
high precision results on \GBee and \GBmumu and
use the \Gll value to check the analysis consistency only.

The extraction of resonance parameters from the measured cross section
requires the accurate accounting of radiative corrections. The Sec.~8.2.4
of the highly cited report~\cite{brambilla-2005} treats the radiative
corrections to $e^{+}e^{-} \to J/\psi \to \ell^{+}\ell^{-}$ cross section
in the way contradicting to that used in the experiments
~\cite{collaboration-2004-69,adams-2006-73} and our
work~\cite{Baldin:2008zz} therefore
me start with the discussion of this issue.

\section{Radiative corrections to $J/\psi$ production and decays}

In virtually all experimental analyses it is assumed that
the resonant contribution to the  cross section of
$e^{+}e^{-} \to J/\psi \to \ell^{+}\ell^{-}$
is proportional to the product \GeeBll where \Gee and \Gll are so called
experimental partial widths~\cite{Tsai1983pr} recommended to use by the
Particle Data Group since 1990:
\begin{equation}
  \Gll \equiv \mathcal{B}_{ll(n\gamma)} \times \Gamma = 
   \frac{\Gamma_{\ell\ell}^{(0)}}{|1-\Pi_0|^2},
\end{equation}
where $\mathcal{B}_{ll(n\gamma)}$ is the branching ratio as it
is measuring experimentally, $\Gamma^{0}_{ee}$ is the lowest order
QED partial width and $\Pi_0$ is the vacuum polarization operator
excluding $J/\psi$ contribution. In contrast, the Sec.~8.2.4
of Ref.~\cite{brambilla-2005} proposes that the resonant contribution
is proportional to 
$\Gamma_{ee}\times\Gamma^{(0)}_{\ell\ell}/\,\Gamma =
\Gamma^{0}_{ee}\times\Gamma_{\ell\ell}/\,\Gamma$.

According to Ref.~\cite{KuraevFadin} the cross section of the
single--photon annihilation $e^+e^- \to \ell^{+}\ell^{-}$ 
can be written in the form
\begin{equation}\label{eq:RadCorInt}
  \sigma = \int\!dx\,
     \frac{\sigma_0((1\!-\!x)s)}{|1-\Pi((1\!-\!x)s)|^2}\, f(s,x),
\end{equation}
where the $f(s,x)$ is calculated with a high accuracy,
the $\Pi(s)$ represents the vacuum polarization operator
and $\sigma_0(s)$ in the Born level cross section of
the process.

 Assuming the Breit-Wigner shape for $\sigma_0$
\begin{equation}
    \sigma(s) = 
\frac{12\pi\,\Gamma^0_{ee}\,\Gamma^0_{\ell\ell}}{(s-M^2)^2+M^2\Gamma^2}
\end{equation}
and replacing $\Pi(s)$ with $\Pi_0$ mentioned above, one reproduces
the result of the Sec.~8.2.4 of Ref.~\cite{brambilla-2005}.

 However,
the Born level cross section of the $e^+e^- \to \ell^{+}\ell^{-}$ 
process is the 
smooth function of $s$ therefore the resonance behavior of the
cross section \eqref{eq:RadCorInt} is due to energy dependence
of the full vacuum polarization operator $\Pi$ containing the resonant
contribution\footnote{We are grateful to V.\,S.~Fadin for clarification of this issue.}.
One has $\Pi = \Pi_0 + \Pi_{R}$ with nonresonant
$\Pi_0 = \Pi_{ee} \!+\! \Pi_{\mu\mu} \!+\! \Pi_{\tau\tau} \!+\!\Pi_{q\bar{q}}$
and
\begin{equation}
\Pi_R(s) = \frac{3\Gamma^0_{ee}}{\alpha}\frac{s}{M_0}
       \frac{1}{s-M_0^2+i M_0\Gamma_0} \,,
\end{equation}
where $M_0$, $\Gamma_0$ and $\Gamma^{(0)}_{ee}$
are the ``bare'' resonance mass and widths. 

 The formula \eqref{eq:RadCorInt} gives the
cross section without separation to
the continuum, resonant and interference parts.
To obtain the contribution of the resonance,
the continuum one must be subtracted from the amplitude. It can be done
with the equality
\begin{equation}\label{eq:Equiv}
\begin{split}
  & \frac{1}{1\!-\!\Pi_0\!-\!\Pi_R(s)} \equiv \\
  & \:\:\:\:\:\:\:\:\:\:\:\:\:\:\:\:\:\:\:\:
  \frac{1}{1\!-\!\Pi_0} +
  \frac{1}{(1\!-\!\Pi_0)^2}\,
  \frac{3\Gamma^0_{ee}}{\alpha}\,\frac{s}{M_0}\,
  \frac{1}{s\!-\!\tilde{M}^2+i \tilde{M}\tilde{\Gamma}}
\end{split}
\end{equation}
in which both $\tilde{M}$ and $\tilde{\Gamma}$ depend on $s$:
\begin{equation}
\begin{split}
 & \tilde{M}^2 = M_0^2 + \frac{3\Gamma^0_{ee}}{\alpha}\,\frac{s}{M_0}\:
                \Re\frac{1}{1-\Pi_0} \,, \\
 & \tilde{M}\tilde{\Gamma} = M_0\Gamma_0 -
     \frac{3\Gamma^0_{ee}}{\alpha}\,\frac{s}{M_0}\: \Im\frac{1}{1-\Pi_0} \,.
\end{split}
\end{equation}
In a vicinity of a narrow resonance this dependence is negligible thus
the resonant contribution can be described with the Breit-Wigner amplitude
containing ``dressed'' parameters $M\approx\tilde{M}(M_0^2)$,
$\Gamma\approx\tilde{\Gamma}(M_0^2)$. Due to the extra power
of the vacuum polarization factor $1/|1-\Pi_0|$ in the second
term of \eqref{eq:Equiv} the resonant part of the 
$e^+e^- \to \ell^{+}\ell^{-}$ cross section is proportional to \GeeBll and
does not depend on $\Gamma^{(0)}_{ee}$ explicitly.

The analytical expressions for the $e^+e^- \to \ell^+\ell^-$ cross section
in the soft photon 
approximation were first derived by  Ya.\,A.~Azimov et al. 
in 1975~\cite{azimov-1975-eng}.
With some up-today modifications one obtains
in the vicinity of a narrow resonance
\begin{equation}
  \label{eq:ee2mumu}
  \begin{aligned}
    &\left(\frac{d\sigma}{d\Omega}\right)^{ee\to\mu\mu}\approx
    \left(\frac{d\sigma}{d\Omega}\right)_{\text{QED}}^{ee\to\mu\mu}+
    \frac{3}{4M^2}
    \left(1+\delta_{\text{sf}}\right)
    \left(1+\cos^2\theta\right) 
    \,\times\\
    &\quad\qquad\left\{
        \frac{3\Gamma_{ee}\Gamma_{\mu\mu}}{\Gamma M}              
        \Im \mathcal{F}- 
        \frac{2\alpha\sqrt{\Gamma_{ee}\Gamma_{\mu\mu}}}{M}\,
        \Re \frac{\mathcal{F}}{1-\Pi_0}
     \right\} ,
  \end{aligned}
\end{equation}
where a correction $\delta_{\text{sf}}$ follows from the 
structure function approach
of~\cite{KuraevFadin}:
\begin{equation}\label{eq:deltasf}
  \delta_{\text{sf}}=\frac{3}{4}\beta+
   \frac{\alpha}{\pi}\left(\frac{\pi^2}{3}-\frac{1}{2}\right)+
  \beta^2\left(\frac{37}{96}-\frac{\pi^2}{12}-
  \frac{1}{36}\ln\frac{W}{m_e} \right)
\end{equation}
and 
\begin{equation}  \label{eq:F}
  \mathcal{F}=\frac{\pi\beta}{\sin\pi\beta}\,
      \left(\frac{M/2}{-W+M-i\Gamma/2}\right)^{1-\beta}
\end{equation}
with 
\begin{equation}
\beta=\frac{4\alpha}{\pi}\left(\ln\frac{W}{m_e}-\frac{1}{2}\right).
\end{equation}
The terms proportional to $\Im\mathcal{F}$ and
$\Re\mathcal{F}$ describe the contribution of the resonance
and the interference effect, respectively.

Originally in Ref.~\cite{azimov-1975-eng} the electron loops only were
taken into account in $\Pi_0$ while
the terms $\lesssim \beta^2$ were omitted
including the $\pi\beta/\!\sin\pi\beta$ factor~\cite{Todyshev} in~\eqref{eq:F}. 

For the $e^{+}e^{-}$ final state one has
\begin{equation}
  \label{eq:ee2ee}
  \begin{aligned}
    &\left(\frac{d\sigma}{d\Omega}\right)^{ee\to ee} \approx 
    \left(\frac{d\sigma}{d\Omega}\right)_{\text{QED}}^{ee\to ee}+\\
    &\quad\frac{1}{M^2}
    \left\{\,\frac{9}{4}\frac{\Gamma^2_{ee}}{\Gamma
      M}(1+\cos^2\theta) \,\left(1+\delta_{\text{sf}}\right)\,\Im\mathcal{F} -\right.\\
    &\qquad\left.\frac{3\alpha}{2}\frac{\Gamma_{ee}}{M}
    \left [(1+\cos^2\theta)-
      \frac{(1+\cos\theta)^2}{(1-\cos\theta)}\right ]
              \Re \mathcal{F}
    \right\},
  \end{aligned}
\end{equation}
where the relative accuracy of the interference term is about $\beta$
(7.6\% for $J/\psi$). That is sufficient for the analysis reported.

For the nonresonant contribution \(\sigma_{\text{QED}}\)  
the calculations of~\cite{Beenakker:1990mb,arbuzov-1997-9710} can be used
implemented in the event generators BHWIDE~\cite{BHWIDE}
and MCGPJ~\cite{MCGPJ}.

In order to compare the theoretical cross
sections~\eqref{eq:ee2mumu} and~\eqref{eq:ee2ee} with experimental data,
it is necessary to perform their convolution with a distribution of the total
collision energy which is
assumed to be Gaussian with an energy spread \(\sigma_W\):
\[\rho(W)=\frac{1}{\sqrt{2\pi}\,\sigma_W}\exp{\left(-\frac{(W-W_0)^2}{2\sigma_W^2}\right)}\,,\]
where \(W_0\) is an average c.m. collision energy. 


\label{sec:VEPP}

\section{VEPP-4M collider and KEDR detector}
\label{sec:VEPP}

The VEPP-4M collider~\cite{Anashin:1998sj} can operate in the wide
range of beam energy from 1 to 6 GeV. The peak luminosity in the
\(J/\psi\) energy region is
about~\(2\times10^{30}\,\text{cm}^{-2}\text{s}^{-1}\).

\begin{center}
  \includegraphics[width=\columnwidth]{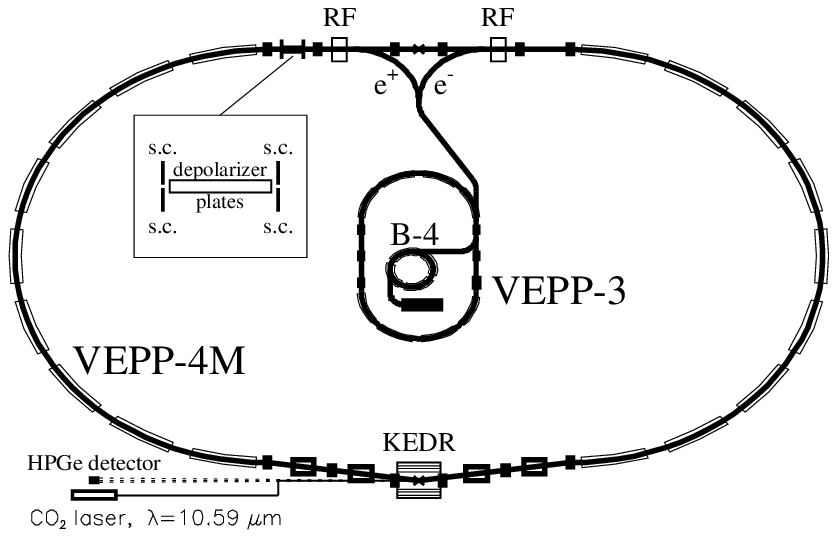}
  \figcaption{\label{fig:vepp4m}
       VEPP-4M/KEDR complex with the resonant depolarization and
    the infrared light Compton backscattering facilities.}
\end{center}

One of the main features of the VEPP-4M is a possibility of precise
energy determination. The resonant depolarization
method~\cite{Bukin:1975db,Skrinsky:1989ie} was implemented at VEPP-4
from the beginning of experiments in early eighties for the
measurements of the \(J/\psi\) and \(\psi(2S)\) mass with the
OLYA~\cite{Artamonov:2000cz} detector and \(\Upsilon\) family mass
with the MD-1~\cite{Artamonov:2000cz} detector.

At VEPP-4M the accuracy of the  energy calibration with the
resonant depolarization is improved to about \(10^{-6}\). The
interpolation of energy between calibrations~\cite{Aulchenko:2003qq}
 in the \(J/\psi\) region has  the accuracy of  \(6\cdot10^{-6}\) ($\simeq$10~keV).

In 2005 a new technique developed at the BESSY-I and BESSY-II
synchrotron radiation sources~\cite{Klein:1997wq,Klein:2002ky} was
adopted for VEPP-4M. It employs the infrared light Compton
backscattering and has a worse precision (50$\div$70~keV in the
\(J/\psi\) region) but, unlike the resonant depolarization, can be
used during data taking.

The KEDR detector~\cite{Anashin:2002uj} includes the vertex detector,
the drift chamber, the scintillation time-of-flight counters, the
aerogel Cherenkov counters, the
barrel liquid krypton calorimeter, the endcap CsI calorimeter, and the
muon system built in the yoke of a superconducting coil generating a
field of 0.65 T. The detector also includes the scattered electron
tagging system for studying of the two-photon processes. The
on-line luminosity is measured by two independent
single bremsstrahlung monitors.

\section{Experiment description}
\label{sec:Exp}

In April 2005, the 11-point scan of the \(J/\psi\) has been performed with
the integral luminosity of 230 nb\(^{-1}\). This
corresponds approximately to 15000 \(J/\psi\to e^+e^-\) decays. During
this time, 26 calibrations of the beam energy were done using the
resonance-depolarization method.

Single bremsstrahlung and Bhabha scattering to the endcap calorimeter
were used in the relative measurement of luminosity. 
The absolute calibration of the luminosity was performed
using the large angle Bhabha scattering in the \GBee analysis.

\begin{center}
  \includegraphics[width=\columnwidth]{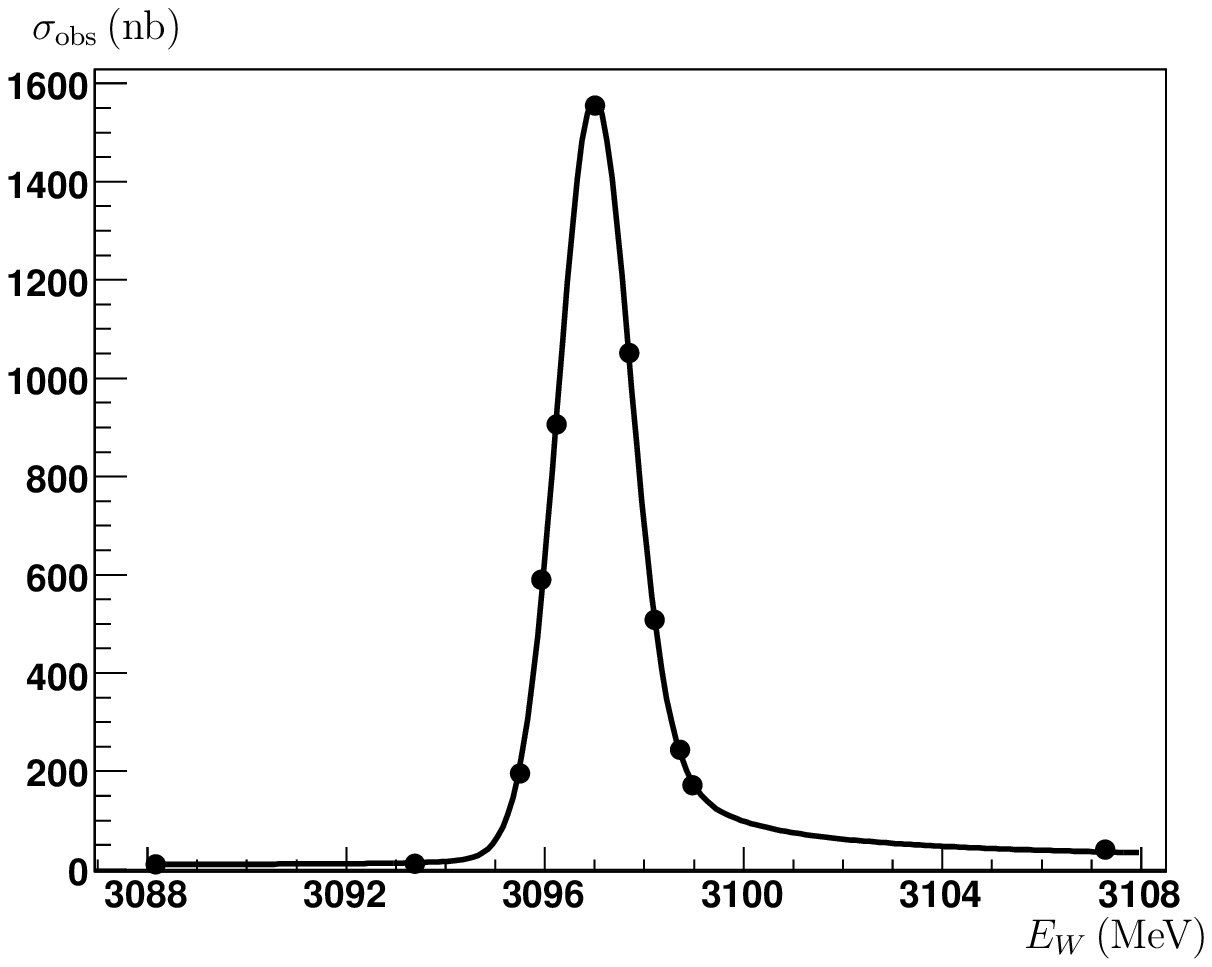}
  \figcaption{\label{fig:mhadr}
      Observed \(e^+e^-\to\text{hadrons}\) cross section
    according to the results  of the \(J/\psi\) scan.}
\end{center}

Figure~\ref{fig:mhadr} shows the observed \(e^+e^-\to\text{hadrons}\)
cross section at the \(J/\psi\) energy region.  These data were used
to fix the resonance peak position and to determine the beam energy
spread.  The value of the \(J/\psi\) mass agrees with the earlier
VEPP-4M/KEDR experiments~\cite{Aulchenko:2003qq}. 

\section{Data analysis}
\label{sec:Data}

In our analysis we employed the simplest selection criteria that
ensured a sufficient suppression of multihadron events and the
cosmic-ray background, please see Ref.~\cite{Baldin:2009pc} for details.

In order to measure the resonance parameters in the $e^{+}e^{-}$ channel,
the set of events was
divided into ten equal angular intervals from \(40^{\circ}\) to
\(140^{\circ}\).  At the $i$-th point in energy \(E_i\) and the
$j$-th angular interval \(\theta_j\), the expected number of
events was parameterized as
\begin{equation}\label{eq:real2sim}
  \begin{aligned}
    N_{\text{exp}}(E_i,\theta_j)=&\mathcal{R}_{\mathcal{L}}\times
    \mathcal{L}(E_i)\times\\
    \Big(&\sigma^{\text{theor}}_{\text{res}}(E_i,\theta_j)\cdot
    \varepsilon^{\text{sim}}_{\text{res}}(E_i,\theta_j)+ \\
    &\sigma^{\text{theor}}_{\text{inter}}(E_i,\theta_j)\cdot
    \varepsilon^{\text{sim}}_{\text{inter}}(E_i,\theta_j)+\\
    &\sigma^{\text{sim}}_{\text{Bhabha}}(E_i,\theta_j)\cdot
    \varepsilon^{\text{sim}}_{\text{Bhabha}}(E_i,\theta_j)
    \Big).
  \end{aligned}
\end{equation}
where \(\mathcal{L}(E_i)\) is the integrated luminosity   measured by
luminosity monitor at
the $i$-th point; \(\sigma^{\text{theor}}_{\text{res}}\),
\(\sigma^{\text{theor}}_{\text{inter}}\) and
\(\sigma^{\text{theor}}_{\text{Bhabha}}\) are the theoretical
cross sections respectively for resonance, interference and Bhabha contributions; 
\(\varepsilon^{\text{sim}}_{\text{res}}\),
\(\varepsilon^{\text{sim}}_{\text{inter}}\) and
\(\varepsilon^{\text{sim}}_{\text{Bhabha}}\) are detector efficiencies
 obtained from simulated data.
 
 In this formula the following free parameters were used:
\begin{enumerate}
\item the product \GBee, which determines the magnitude of the
  resonance signal;
\item the electron width \(\Gamma_{ee}\), which specifies the
  amplitude of the interference wave;
\item the coefficient \(\mathcal{R}_{\mathcal{L}}\), which provides
  the absolute calibration of the luminosity monitor.
\end{enumerate}

\begin{center}
 \includegraphics[width=\columnwidth]{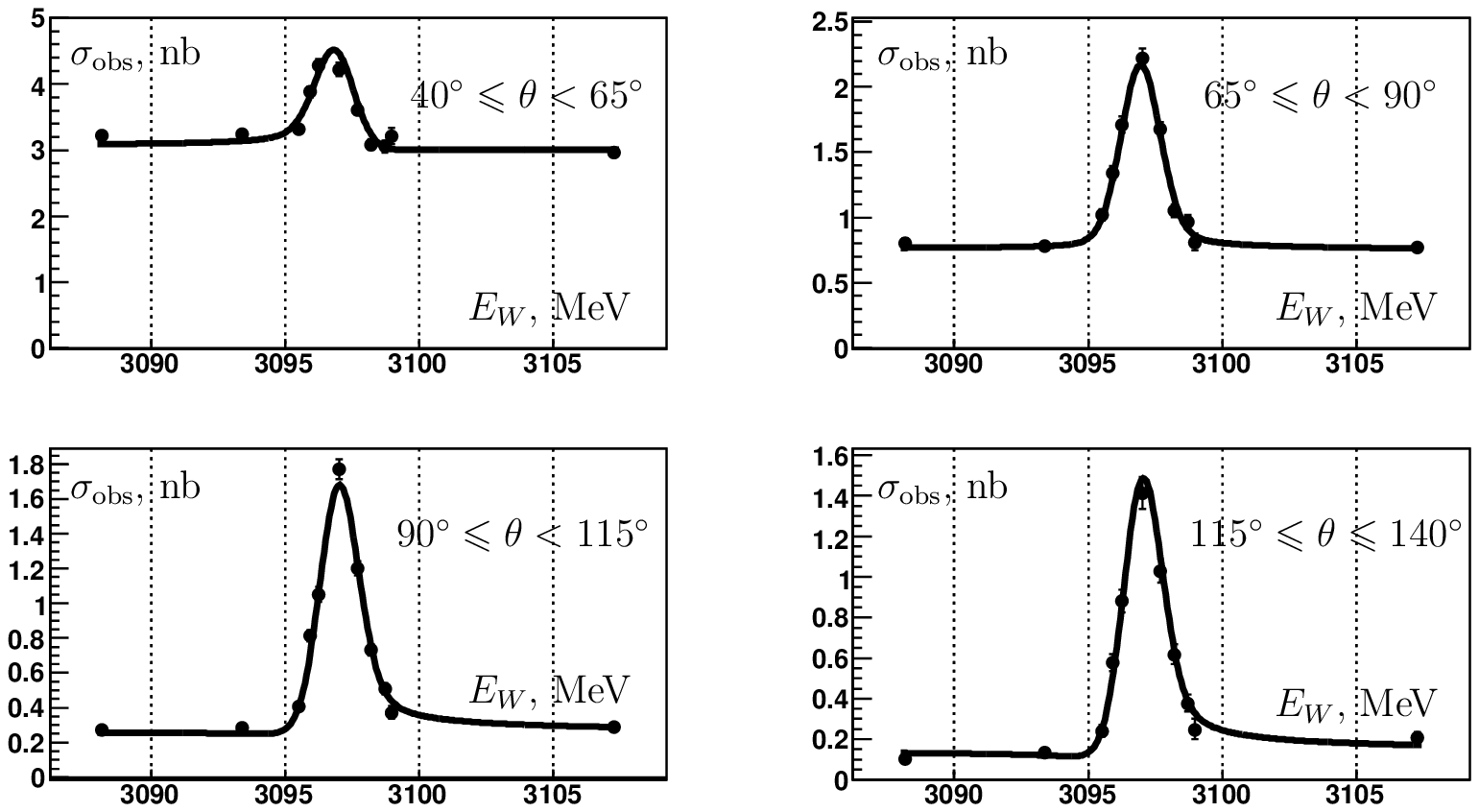}
  \figcaption{\label{fig:real2sim}
    Fits to experimental data for $e^+e^-\to e^+e^-$ process
    at $J/\psi$ energy region for four angular ranges.}  
\end{center}

We note that the coefficient \(\mathcal{R}_{\mathcal{L}}\) partially 
accounts the possible difference between the actual detection efficiency
and simulation in the case where these difference do not depend on the
scattering angle or the beam energy (or the data taking time)
thus the substantial cancellation of errors occurs. 

Figure~\ref{fig:real2sim} shows our fit to the data for four angular
intervals.  
The joined fit in ten equal intervals from \(40^\circ\) to
\(140^\circ\) produce the following
basic result:
\begin{equation}\label{eq:eeresult}
  \begin{split}
    &\GBee = 0.3323\pm0.0064\,\text{(stat.)}\,\text{keV},\\
    &\mathcal{R}_{\mathcal{L}}=93.4\pm0.7\,\text{(stat.)}\,\%,\\
    &\Gamma_{ee} = 5.7 \pm0.6\,\text{(stat.)}\,\text{keV}.
 \end{split}
\end{equation}

Due to different angular distributions for Bhabha scattering and
resonance events, subdivision of the data into several angular bins decreases
a statistical error for \GBee by \(40\div50\,\%\).
The electron width obtained by the fit has a statistical error of
about 10\% and agrees with the world-average value.

\begin{center}
  \includegraphics[width=\columnwidth]{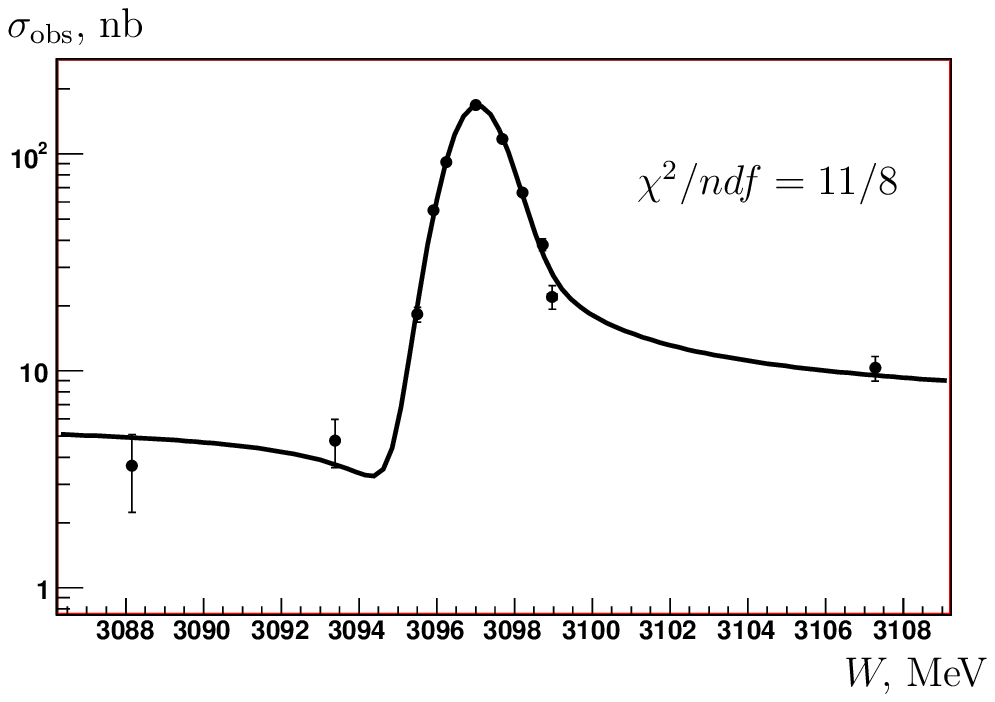}
  \figcaption{\label{fig:mumufit}
    Fit to experimental data for $e^+e^-\to \mu^+\mu^-$ process
    at $J/\psi$ energy region.}
\end{center}

Similarly to~\eqref{eq:real2sim}, the expected number of $e^+e^-\to
\mu^+\mu^-$ events was parameterized in the form:
\begin{equation}
  \begin{aligned}
    N_{\text{exp}}(E_i)=&\mathcal{R}_{\mathcal{L}}\times
       \mathcal{L}(E_i)\times\\
    \Big(&\sigma^{\text{theor}}_{\text{res}}(E_i)\cdot
      \varepsilon^{\text{sim}}_{\text{res}}(E_i)+\\
    &\sigma^{\text{theor}}_{\text{inter}}(E_i)
     \cdot\varepsilon^{\text{sim}}_{\text{inter}}(E_i)+\\
    &\sigma^{\text{theor}}_{\text{bg}}(E_i)
     \cdot\varepsilon^{\text{sim}}_{\text{bg}}(E_i)
    \Big)+F_{\text{cosmic}}\times T_i,
  \end{aligned}\label{eq:mumufit}
\end{equation}
with the same meaning of \(\mathcal{R}_{\mathcal{L}}\) and
\(\mathcal{L}(E_i)\) as  in~\eqref{eq:real2sim}. 
\(\mathcal{L}(E_i)\)  is multiplied
by the sum of the products of theoretical cross sections for
resonance, interference and QED background and detection efficiencies
as obtained from simulated data.  
\(\mathcal{R}_{\mathcal{L}}\) 
was fixed by result~\eqref{eq:eeresult}.  \(T_i\)
is the live data taking time. Unlike~\eqref{eq:real2sim} there is 
only one angular interval  from \(40^{\circ}\) to \(140^{\circ}\).

The following free parameters were used:  
\begin{enumerate}
\item the product \GBmumu, which determines the magnitude of the
  resonance signal;
\item the square root of electron and muon widths
  \(\sqrt{\Gamma_{ee}\Gamma_{\mu\mu}}\), which specifies the amplitude
  of the interference wave;
\item the cosmic events rate \(F_{\text{cosmic}}\) passed
  the selection criteria for the \(e^+e^-\to\mu^+\mu^-\) events.
\end{enumerate}
Due to variations of luminosity during the experiment it is possible
to separate cosmic events contribution (\(F_{\text{cosmic}}\cdot T_i\))
from nonresonant background contribution
(\(\sigma^{\text{theor}}_{\text{bg}}(E_i)\cdot\varepsilon^{\text{sim}}_{\text{bg}}(E_i)\cdot L_i\)).

Figure~\ref{fig:mumufit} shows our fit to the \(e^+e^-\to\mu^+\mu^-\)
data.  It yields the following  result:
\begin{equation}\label{eq:mmresult}
  \begin{split}
    &\GBmumu = 0.3318\pm0.0052\,\text{(stat.)}\,\text{keV},\\
    &\sqrt{\Gamma_{ee}\times\Gamma_{\mu\mu}}=5.6\pm0.7\,\text{(stat.)}\,\text{keV}.
  \end{split}
\end{equation}
As can be seen from~\eqref{eq:mmresult} the statistical error of 
\GBmumu is about 1.6\%.

\section{Discussion of systematic uncertainty}

\label{sec:ErrDisc}

The most significant systematic uncertainties in the \GBee and 
\GBmumu measurements are
listed in Tables~\ref{tab:ee:systematic} and~\ref{tab:mumu:systematic},
respectively.

\begin{center}
\tabcaption{\label{tab:ee:systematic} Systematic uncertainties in \GBee.}
\footnotesize
\medskip
\begin{tabular*}{80mm}{lc}
\toprule
Systematic uncertainty source & \hspace*{-1.25em}Error\,\%\\
\hline
Luminosity monitor instability & 0.8\\
Offline event selection & 0.7\\
Trigger efficiency      & 0.5\\
Energy spread accuracy &0.2\\
Beam energy measurement (10--30\,keV) & 0.3\\
Fiducial volume cut &0.2\\
Calculation of radiative correction & 0.2\\
Cross section for Bhabha  (MC generators) & 0.4\\
Uncertainty in the final state radiation (PHOTOS)  & 0.4\\
Background from  \(J/\psi\) decays&  0.2\\
Fitting procedure & 0.2\\
\hline
\textit{Quadratic sum}&\textit{1.4}\\
\bottomrule
\end{tabular*}
\end{center}

\begin{center}
\tabcaption{\label{tab:mumu:systematic}Systematic uncertainties in \GBmumu.}
\footnotesize
\begin{tabular*}{80mm}{lc}
\toprule
Systematic uncertainty source & \hspace*{-1.25em}Error\,\%\\
\hline
Luminosity monitor instability  & 0.8\\
Absolute luminosity calibration by $e^+e^-$ data &1.2\\
Trigger efficiency      & 0.5\\
Energy spread accuracy &0.4\\
Beam energy measurement (10--30\,keV) & 0.5\\
Fiducial volume cut &0.2\\
Calculation of radiative correction & 0.1\\
Uncertainty in the final state radiation (PHOTOS)  & 0.5\\
Nonresonant background  & 0.1\\
Background from  \(J/\psi\) decays &  0.6\\
\hline
\textit{Quadratic sum}&\textit{1.8}\\
\bottomrule
\end{tabular*}
\end{center}

A rather large  uncertainty of 0.8\% common for the electron and muon
channels is due to the luminosity monitor instability.
It was estimated from comparing the results obtained using the
on-line luminosity of the single bremsstrahlung monitor and
the off-line luminosity
measured by the $e^+e^-$ scattering in the endcap
calorimeter.

The essential source of uncertainty is an imperfection of the detector
response simulation resulting in the errors in
the trigger and  offline event selection efficiencies.
It was studied using collected data and the correction
of $0.75\pm0.7$\% was applied.

The dominant uncertainty of the \GBmumu result is
associated with the absolute luminosity calibration done
in $e^+e^-$-channel.
It includes the accuracy of the Bhabha event generators,
the statistic error of $\mathcal{R}_\mathcal{L}$
parameter~\eqref{eq:eeresult} and the residual (after correction using
simulated data) efficiency difference
for $e^+e^-$ and $\mu^+\mu^-$ events. The additional correction applied to
this difference is  $-0.5\pm0.9$\%.

The other sources of uncertainty are discussed in Ref.~\cite{Baldin:2009pc}.

\section{Results and Conclusion}

The new measurement of the \GBee and \GBmumu has been performed at the
VEPP-4M collider using the KEDR detector.  The following results have
been obtained (in keV):
\begin{equation*}
  \begin{split}
    &\GBee\,=0.3323\pm0.0064\,\text{(stat.)}\,\pm0.0048\,\text{(syst.)}\\
    &\GBmumu=0.3318\pm0.0052\,\text{(stat.)}\,\pm0.0063\,\text{(syst.)}
  \end{split}
\end{equation*}

\begin{center}
  \includegraphics[width=\columnwidth]{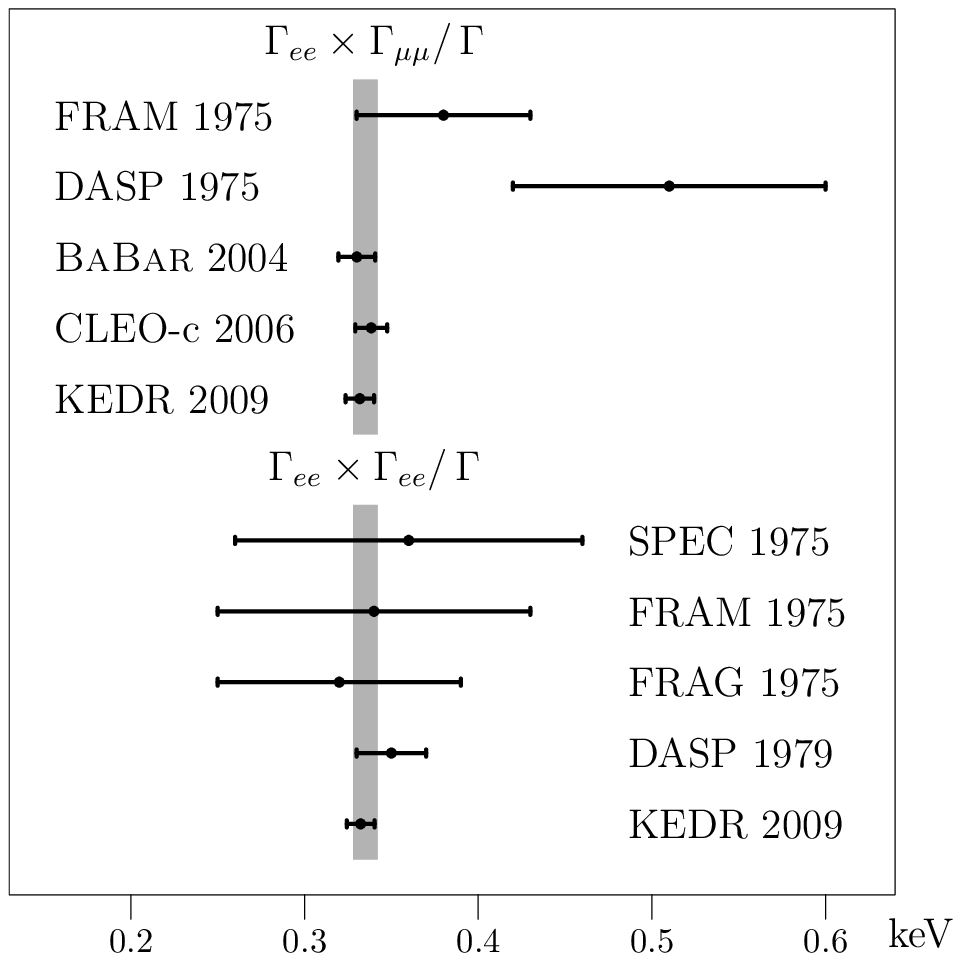}
  \figcaption{\label{fig:cchist}
    Comparison of \GBee and \GBmumu measured at different
    experiments mentioned in~\cite{PDG-2008} with KEDR 2009 results.
    The vertical strip is for the world average \GBmumu value.}
\end{center}

Figure~\ref{fig:cchist} shows the comparison of our results with those of the
previous experiments.
The grey line shows PDG average and the error for the \GBmumu product
measurement. The new KEDR results are the most precise. Results are in
good agreement with each other and with the world average value of
\GBmumu.

Accounting the correlations in the \GBee and \GBmumu errors 
the mean value is
\begin{equation*}
    \GeeBll =0.3320\pm0.0041\,\text{(stat.)}\,\pm0.0050\,\text{(syst.)}
\,\,\text{keV.} 
\end{equation*}

With the assumption of lepton universality and using independent
data on branching fraction 
\(\mathcal{B}(J/\psi\to e^+e^-)=(5.94\pm0.06)\,\%\)~\cite{PDG-2008} leptonic and full widths
of \(J/\psi\) meson were determined:
\begin{equation*}
  \label{eq:result:Gll}
  \begin{split}
    &\Gamma_{\ell\ell}=5.59\pm0.12\,\,\text{keV}\\ 
    &\Gamma\,=\,94.1\pm2.7\,\,\text{keV} 
  \end{split}
\end{equation*}
These results are in good agreement with the world average~\cite{PDG-2008}
and with the results from
\textsc{BaBar}~\cite{collaboration-2004-69} and
\mbox{CLEO-c}~\cite{adams-2006-73} experiments.
 
\end{multicols}

\vspace{-2mm}
\centerline{\rule{80mm}{0.1pt}}
\vspace{2mm}

\begin{multicols}{2}

\end{multicols}

\end{document}